\documentclass[structabstract]{aa}  
\usepackage{graphicx} 
\usepackage{txfonts}
\usepackage{natbib}
\usepackage{hyperref}
\usepackage{siunitx}  

\newcommand{\GG}{\mbox{$G$}}
\newcommand{\GBP}{\mbox{$G_{\rm BP}$}}
\newcommand{\GRP}{\mbox{$G_{\rm RP}$}}

\newcommand{\nueff}{\mbox{$\nu_{\rm eff}$}}
\newcommand{\Zthree}{\mbox{$Z_{\rm EDR3}$}}
\newcommand{\Zfive}{\mbox{$Z_{5}(G,\nu_{\rm eff},\beta)$}}
\newcommand{\Zsix}{\mbox{$Z_{6}(G,\nu_{\rm eff},\beta)$}}
\newcommand{\kfive}{\mbox{$k_{5}$}}
\newcommand{\angcov}{\mbox{$V_\varpi$}}
\newcommand{\sigmas}{\mbox{$\sigma_{\rm s}$}}
\newcommand{\sigmai}{\mbox{$\sigma_{\rm int}$}}
\newcommand{\sigmae}{\mbox{$\sigma_{\rm ext}$}}
\newcommand{\sigman}{\mbox{$\sigma_{\rm n}$}}

\newcommand{\microas}{\mbox{$\mu$as}}
\newcommand{\microninv}{\mbox{$\mu$m$^{-1}$}}
\newcommand{\mci}[1]{\multicolumn{1}{c}{#1}}
\newcommand{\mcii}[1]{\multicolumn{2}{c}{#1}}

\newcommand{\pmra}{\mbox{$\mu_{\alpha *}$}}
\newcommand{\pmdec}{\mbox{$\mu_{\delta}$}}
\newcommand{\pmrag}{\mbox{$\mu_{\alpha *,{\rm g}}$ }}
\newcommand{\pmdecg}{\mbox{$\mu_{\delta,{\rm g}}$}}
\newcommand{\pic}{\mbox{$\varpi_{\rm c}$}}

\newcommand{\pig}{\mbox{$\varpi_{\rm g}$}}
\newcommand{\spig}{\mbox{$\sigma_{\rm g}$}}
\newcommand{\pin}{\mbox{$\varpi_{\rm n}$}}
\newcommand{\tpi}{\mbox{$t_\varpi$}}
\newcommand{\tra}{\mbox{$t_{\mu_{\alpha *}}$}}
\newcommand{\tdec}{\mbox{$t_{\mu_{\delta}}$}}

\begin{document}

   \title{Validation of the accuracy and precision of \textit{Gaia} EDR3 parallaxes with globular clusters}
   \titlerunning{Accuracy and precision of \textit{Gaia} EDR3 parallaxes}

   \author{J. Ma\'{\i}z Apell\'aniz \inst{1}
           \and
           M. Pantaleoni Gonz\'alez\inst{1,2}
           \and
           R. H. Barb\'a\inst{3}
           }
   \authorrunning{J. Ma\'{\i}z Apell\'aniz et al.}

   \institute{Centro de Astrobiolog\'{\i}a, CSIC-INTA. Campus ESAC. 
              C. bajo del castillo s/n. 
              E-\num{28692} Villanueva de la Ca\~nada, Madrid, Spain\linebreak
              \email{jmaiz@cab.inta-csic.es} \\
              \and
              Departamento de Astrof{\'\i}sica y F{\'\i}sica de la Atm\'osfera, Universidad Complutense de Madrid. 
              E-\num{28040} Madrid, Spain. \\
              \and
              Departamento de Astronom{\'\i}a, Universidad de La Serena.
              Av. Cisternas 1200 Norte.
              La Serena, Chile \\
             }

   \date{Received 25 January 2021 / Accepted 3 March 2021}

 
  \abstract
   {The recent early third data release (EDR3) from the \textit{Gaia} mission has produced parallaxes for $1.468 \times 10^9$ sources with better quality
    than for the previous data release. Nevertheless, there are calibration issues with the data that require corrections to the published values and 
    uncertainties.}
   {We want to properly characterize the behavior of the random and systematic uncertainties of the \textit{Gaia}~EDR3 parallaxes in order to maximize the 
    precision of the derived distances without compromising their accuracy. We also aim to provide a step-by-step procedure for the calculation of distances
    to stars and stellar clusters when using such parallaxes.}
   {We reanalyze some of the data presented in the calibration papers for quasar and Large Magellanic Cloud (LMC) parallaxes and combine those results with 
    measurements for six bright globular clusters. We calculate the angular covariance of EDR3 parallaxes at small separations (up to a few degrees) based on the LMC 
    results and combine it with the results for larger angles using quasars to obtain an approximate analytical formula for the angular covariance over the whole sky. The 
    results for the six globular clusters are used to validate the parallax bias correction as a function of magnitude, color, and ecliptic latitude derived
    by Lindegren et al. and to determine the multiplicative constant $k$ used to convert internal uncertainties to external ones.}
   {The angular covariance at zero separation is estimated to be 
    106~$\microas^2$, 
    yielding a minimum (systematic) uncertainty for EDR3 parallaxes of 
    10.3~\microas\ for individual stars or compact stellar clusters. That value can be only slightly reduced for globular clusters that span 
    $\gtrsim$30\arcmin\ after considering the behavior of the angular covariance of the parallaxes for small separations. The Lindegren et al. parallax bias 
    correction is found to work quite well, except perhaps for the brighter magnitudes, where improvements may be possible.
    The value of $k$ is found to be 1.1-1.7 and to depend on \GG. We find that stars with moderately large values of the Renormalized
    Unit Weight Error (RUWE) can still provide useful parallaxes albeit with larger values of $k$. We give accurate and precise \textit{Gaia}~EDR3 distances to the six
    globular clusters and for the specific case of 47~Tuc we are able to beat the angular covariance limit 
    through the use of the background SMC as a reference
    and derive a high-precision distance of 4.53$\pm$0.06~kpc.
    Finally, a recipe for the derivation of distances to stars and stellar clusters using \textit{Gaia}~EDR3 parallaxes is given.
    }
   {}
   \keywords{astrometry -- globular clusters: general -- methods: data analysis -- parallaxes -- stars: distances -- surveys}
   \maketitle
%

\section{Introduction}

$\,\!$\indent The early third {\it Gaia} data release (EDR3, \citealt{Browetal20}) was presented on 3~December~2020 and included photometry for $1.8 \times 10^9$ 
sources and astrometry for $\sim$80\% of them. The ``early'' in EDR3 refers to the fact that the associated spectrophotometry will be presented at a later date. 
The astrometric solution is presented in \citet{Lindetal20a}, from now on L20b, and includes an analysis of the angular covariance of the parallaxes at different
angular scales using data for quasars (large separations $\theta$) and stars in the LMC (small separations). \citet{Lindetal20b}, from now on L20a, analyzes the parallax
bias (or zero point, \Zthree) in the data as a function of magnitude (\GG, the primary very broad band optical photometry provided by {\it Gaia}), 
color (\nueff, the effective wavenumber, which for most good-behaved sources is a function of the \GBP$-$\GRP\ color provided by {\it Gaia}, see Fig.~2 in L20b),
and ecliptic latitude ($\beta$). The median parallax for quasars measured by L20a is $-$17~\microas, which can be understood as the typical parallax bias given
that quasars are too far away to have parallaxes measurable by {\it Gaia}, but systematic variations as a function of position in the sky at the level of 
10~\microas\ are detected (Fig.~2 in L20a). Furthermore, L20a use data from LMC stars and physical binaries to show that \Zthree\ has a complex behavior when 
going from
the (mostly faint) quasar regime to brighter objects with even larger deviations from the median parallax value (see Fig.~20 in that paper), hence the 
need to characterize \Zthree\ at least as a function of \GG, \nueff, and $\beta$.

In this paper we have three objectives: (1) to validate the results from L20a and L20b with an independent dataset built from globular clusters, 
(2) to provide a general recipe for the use of {\it Gaia}~EDR3 parallaxes to derive precise and 
accurate\footnote{We define a result as precise when it has a small total uncertainty and as accurate when its systematic uncertainty is either small
or is properly corrected for. See section~2 for a further description of how we define the different uncertainty types.
}
distances to stars and stellar clusters, and 
(3) to analyze cases where it may be possible to beat the parallax bias to reduce the uncertainty in those distances. First, we present the formalism 
we will use throughout the paper; second, we reevaluate the angular covariance of the {\it Gaia}~EDR3 parallaxes; 
third, we briefly discuss the anchoring of the parallaxes to distant objects; and fourth,
we use a sample of six
globular clusters to validate the \Zthree\ from L20a and estimate the external uncertainties of the parallaxes. We then discuss the possibility of obtaining even
more precise distances for globular clusters using {\it Gaia}~EDR3 data and we end up presenting a summary of the paper and our proposed recipe for the
calculation of distances from {\it Gaia}~EDR3 parallaxes.

\section{Formalism for the analysis of \textit{Gaia} EDR3 parallaxes}

$\,\!$\indent In this section we present the formalism that will be used in this paper and that follows for the most part that of \citet{Lindetal18b}, from now
on L18, plus L20a and L20b. 

\subsection{Correcting the parallax bias}

$\,\!$\indent The relationship between the measured EDR3 parallax, $\varpi$, and the corrected EDR3 parallax, \pic, is given by:

\begin{equation}
 \pic = \varpi - \Zthree,
\label{pic}
\end{equation}

\noindent where, following L20a, \Zthree\ is a function of \GG, \nueff, and $\beta$ and takes different forms for five-parameter solutions as \Zfive\ and for
six-parameter solutions as \Zsix. The separation is needed because {\it Gaia}~EDR3 astrometry comes in two flavors: for those with measured \nueff, five
parameters are fitted (two coordinates, one parallax, and two proper motions), while for those where \nueff\ is not known a priori it has to be added as a sixth
parameter (the pseudocolor). Five-parameter solutions are of better quality and are the majority for stars brighter than \GG~=~19 (Fig.~5 in L20b) but a minority
(40\%) of the total {\it Gaia}~EDR3 sample, which is dominated by sources in the \GG~=~19-21.5 range. As in this paper we are interested mostly in bright stars,
we will pay attention primarily to those with five-parameter solutions though for the globular cluster sample we will also include a small fraction of stars
with six-parameter solutions. \Zfive\ and \Zsix\ are defined from three orthogonal polynomial basis functions $b_0,b_1,b_2$ of the ecliptic latitude (of zeroth, 
first, and 
second order in $\sin\beta$,
respectively), five piecewise basis functions $c_0,c_1,c_2,c_3,c_4$ that depend on \nueff\ (Appendix A in L20a), and  
tabulated $q_{jk}$ coefficients for 13 values of \GG\ between 6 and 21: 

\begin{equation}
 \Zthree(G,\nu_{\rm eff},\beta) = \sum_{j=0}^{4} \sum_{k=0}^{2} q_{jk}(G)\, c_j(\nueff)\, b_k(\beta),
\label{Zdef}
\end{equation}

\noindent where two different tables are given for \Zfive\ and \Zsix\ (Tables~9~and~10 in L20a, respectively). From the tabulated values of $q_{jk}$ one can 
interpolate to any \GG\ magnitude. For \Zfive, only eight out of the possible fifteen combinations of $c_j$ and $b_k$ have non-zero values of $q_{jk}$ and of 
those only four have non-zero values for the whole \GG~=~6-21 range: $q_{00}$, $q_{01}$, $q_{02}$, and $q_{11}$, the three color-independent $\beta$ terms and 
the term that depends linearly on both \nueff\ (in the 1.24-1.48 \microninv\ range) and $\beta$. The remaining four combinations with non zero-terms are 
$q_{10}$, $q_{20}$, $q_{30}$, and $q_{40}$ are all $\beta$-independent and apply to stars 
brighter than \GG~=13.1 ($q_{10}$, linear term in the 1.24-1.48 \microninv\ range), 
brighter than \GG~=~17.5 ($q_{20}$, applicable for stars redder than \nueff~=~1.48~\microninv), 
between \GG~=~13.1~and~17.5 ($q_{30}$, applicable for stars redder than \nueff~=~1.24~\microninv), and
between \GG~=~13.1~and~19.0 ($q_{40}$, applicable for stars bluer than \nueff~=~1.72~\microninv), respectively.

\subsection{Accounting for the uncorrected parallax bias}

$\,\!$\indent The L20a \Zthree$(G,\nu_{\rm eff},\beta)$ parallax-bias corrections improve the consistency of the parallaxes of stars in stellar clusters, as shown
in Figs.~17~and~18 of \citet{Fabretal20} and as we also show later on in this paper. Of course, an empirical correction like that of L20a can always be improved
with better data but for the time being we may assume that it removes most of the bias associated with magnitude and color and with the spherical 
harmonics of degree $\le 2$
of the angular power spectrum of the parallax bias. However, L20b used the parallaxes of 1 million quasars to show that the angular power spectrum of 
the parallax bias is dominated by the contributions of the spherical harmonics of higher order. Putting it in another way, L20b determined that the angular 
covariance \angcov\ of the quasar parallaxes can be approximately described by:

\begin{equation}
 \angcov(\theta) = 142\,\exp(-\theta/16^{\rm o})\, \microas^2
\label{angcov1}
\end{equation}

\noindent for separations $0.5^{\rm o} < \theta < 80^{\,\rm o}$. Taking Eqn.~\ref{angcov1} at face value, the square root of its value at zero separation,
11.9~\microas, represents the dispersion of the uncorrected parallax bias and can be thought of as an additional uncertainty source for any {\it Gaia}~EDR3 
parallax, which we will refer to in this paper as \sigmas. Furthermore, as any star located a short distance away from the source of interest will have a similar 
uncorrected parallax bias, \sigmas\ acts as a systematic effect i.e. it affects all stars in a compact cluster in the same (unknown for a specific object) manner 
and sets a limit on the overall uncertainty of the cluster parallax when combining data from many stars.
However, there are at least two reasons why 
Eqn.~\ref{angcov1} is not the whole story. The first reason is that the quasar sample used by L20b is very faint, with a median \GG\ of 19.9~mag. This raises the 
question of whether the angular spectrum of the parallax bias for brighter sources has the same characteristics, something that is answered below. The 
second reason is that Eqn.~\ref{angcov1} does not reflect the behavior of \angcov\ at small separations (from a fraction of to a few degrees) as, indeed, the 
value of the first bin ($\theta < 0.125^{\rm o}$) for the quasar covariance reported by L20b has a significantly larger value of 700~$\microas^2$ (albeit with a 
large uncertainty) that leads to \sigmas~=~26.5~\microas. That small-angle behavior is better captured by the checkered pattern seen in the smoothed parallaxes 
for LMC sources with \GG~=16-18~mag shown in Fig.~14 of L20b. The pattern has a similar triangular disposition of maxima and minima with peak-to-peak or
valley-to-valley separation of $\sim 1^{\rm o}$ in {\it Gaia}~DR2~and~EDR3 but the RMS amplitude has decreased from 13.1~\microas\ to 6.9~\microas\ between the two
data releases. In the next section we combine these pieces of information to generate an approximate analytical form for \angcov\ for any separation.

A different issue that appears when comparing {\it Gaia} parallaxes with external data is that the (random) internal uncertainties \sigmai\ do not reflect the
true dispersion of the values listed in the catalog. Here we follow L18 to define the total or external uncertainty \sigmae\ for a parallax as:

\begin{equation}
 \sigmae = \sqrt{k^2\sigmai^2 + \sigmas^2}, 
\label{sigmae}  
\end{equation}

\noindent where $k$ is a multiplicative constant that needs to be determined and that may depend on magnitude or other quantities. For a well-characterized
catalog it is expected to be close to one. \citet{Fabretal20}, who call it uwu for unit-weight uncertainty, indeed find that for most of the {\it Gaia}~EDR3 
samples they analyzed it is between 1.0 and 1.8 (see their Table~1 and their Fig.~21). Later in this paper we provide an independent evaluation of $k$.

\subsection{Combining parallaxes}

$\,\!$\indent Using Equation~\ref{sigmae} to obtain the total parallax uncertainty of a source leads into a problem: it is neither a random nor a systematic
uncertainty but a combination of both. 
More specifically, it can be treated as a random uncertainty but only if we are dealing with a single source and calculating its distance but not if we are
dealing with several sources assumed to be at the same distance.
If we want to obtain an improved parallax by combining the individual corrected parallaxes $\varpi_{{\rm c},i}$ from the 
members of a multiple stellar system or cluster, we should treat each type of uncertainty properly. Furthermore, when combining information from stars at 
non-negligible separations we should use \angcov\ to determine how to treat the correlated uncertainties. The answer is to use a slightly modified version of the 
\citet{Campetal19} procedure to obtain the group parallax \pig:

\begin{equation}
 \pig = \sum_{i=1}^{n} w_i \varpi_{{\rm c},i},
\label{pig}  
\end{equation}

\noindent where the weights are given by:

\begin{equation}
 w_i = \frac{1/\sigma_{{\rm ext},i}^2}{\sum_{i=1}^{n} 1/\sigma_{{\rm ext},i}^2}
\label{weights}  
\end{equation}

\noindent and the group parallax uncertainty is given by:

\noindent 
\begin{equation}
 \sigma^2_{\rm g} = \sum_{i=1}^{n} w_i^2\,\sigma_{{\rm ext},i}^2 + 2 \sum_{i=1}^{n-1} \sum_{j=i+1}^{n} w_i\,w_j\,\angcov(\theta_{ij}),
\label{spig}  
\end{equation}

\noindent where the first term reflects the contributions from the individual stars and the second term is a sum over all pairs of stars to properly account for
the correlations introduced by the angular covariance. For the simple case where all external uncertainties are the same and the same applies to the angular
covariance, the first term becomes $\sigma_{{\rm ext},i}^2/n$ and the second term 
becomes $\angcov(\theta_{ij})(n-1)/n$,
that is, the first term makes \spig\ 
improve with the square root of the number of stars while the second one is relatively insensitive to how many objects are in the group. Therefore, in the limit
of a cluster with a large number of stars its parallax uncertainty becomes the square root of the angular covariance averaged over all stellar 
pairs\footnote{It is possible in principle to play with a selection of which stars to use and which ones to discard to minimize even further the result from 
Eqn~\ref{spig}. However, when attempting this process with real stellar clusters the improvement in \spig\ is usually very small, of the order of 10\% at most.}.
This effect is seen in Table~3 of \citet{Maizetal20b}, who used this strategy to obtain the group parallaxes of OB stellar groups from {\it Gaia}~DR2 parallaxes. Most
group uncertainties have values close to 43~\microas, as that paper used a value of $\angcov(0)$ of 1850~$\microas^2$ (L20b and references therein) and most
of the stellar groups have small angular sizes. The two most notable exceptions are Villafranca~O-015 (Collinder~419) and Villafranca~O-016 (NGC~2264), see also
\citet{Maiz19}, as those two stellar groups are nearby and they have larger angular sizes, allowing \angcov\ to be averaged over longer separations and hence become
lower. That effect reduces their group parallax uncertainties to 34~\microas\ and 29~\microas, respectively.

Below we apply this procedure to determine the group parallaxes and their uncertainties for six globular clusters using {\it Gaia}~EDR3 data. The results are then
used to validate the parallax bias and to determine $k$.

\subsection{Calculating distances}

$\,\!$\indent The last step that is usually required to use a parallax is to convert it into a distance $d$ or, more precisely, a posterior distribution of
distances $p(d|\pic)$. To obtain such a distribution for given values of \pic\ and \sigmae\ one needs to use a likelihood distribution $p(\pic|d)$ (a gaussian is
typically assumed) and a prior distribution for the distances to the population to which the object belongs to. It has been known for a long time 
\citep{LutzKelk73} that a flat prior that extends to infinity makes the posterior distribution diverge at large distances. Therefore, a prior that goes to
zero faster than $1/d^3$ is required (see Eqn.~1 in \citealt{Maiz05c}). Given the finite extent of the Milky Way, this should not be a problem when dealing with
Galactic sources but, nevertheless, one should not
apply a prior blindly as the underlying spatial distribution is not the same depending on e.g. whether the star is a red giant or an early-type star or 
whether it belongs to the disk or the halo populations. This issue is even more relevant when a sightline contains objects concentrated at very different
distances such as the solar neighborhood, a globular cluster, and a Local Group galaxy. On the other hand, Pantaleoni Gonz\'alez et al.
(2021, submitted to MNRAS) have compared {\it Gaia}~DR2 distances to OB stars using three different priors, one specific for OB stars 
\citep{Maiz01a,Maizetal08a} and two for the general Galactic (mostly disk) population \citep{Bailetal18,Andeetal19}, and have found out that the results are quite
similar. There are some systematic differences but they are well within the posterior uncertainties. That result means that as long as the prior is a reasonable
description of the underlying population, distances derived from {\it Gaia}~parallaxes are robust and mostly independent of the details of the prior itself.

\section{The angular covariance of the \textit{Gaia} EDR3 parallaxes}

$\,\!$\indent In this section we re-evaluate the angular covariance of the \textit{Gaia} EDR3 parallaxes. We analyze first the small separations case ($\theta < 4^{\rm o}$)
and we then extend the analysis to larger values of $\theta$.

\subsection{The small-separation angular covariance}

\begin{figure*}
 \centerline{\includegraphics[width=0.49\linewidth]{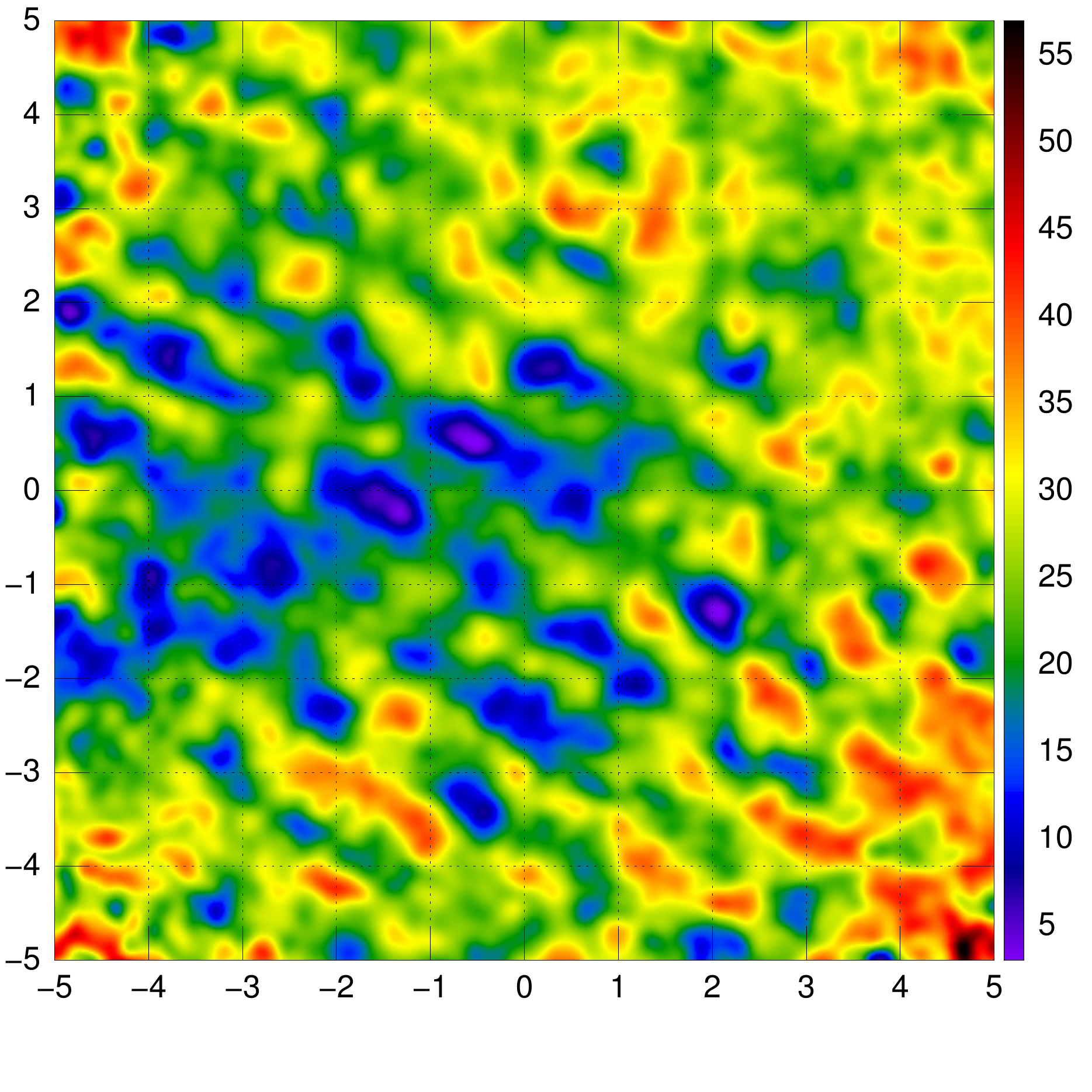} \
             \includegraphics[width=0.49\linewidth]{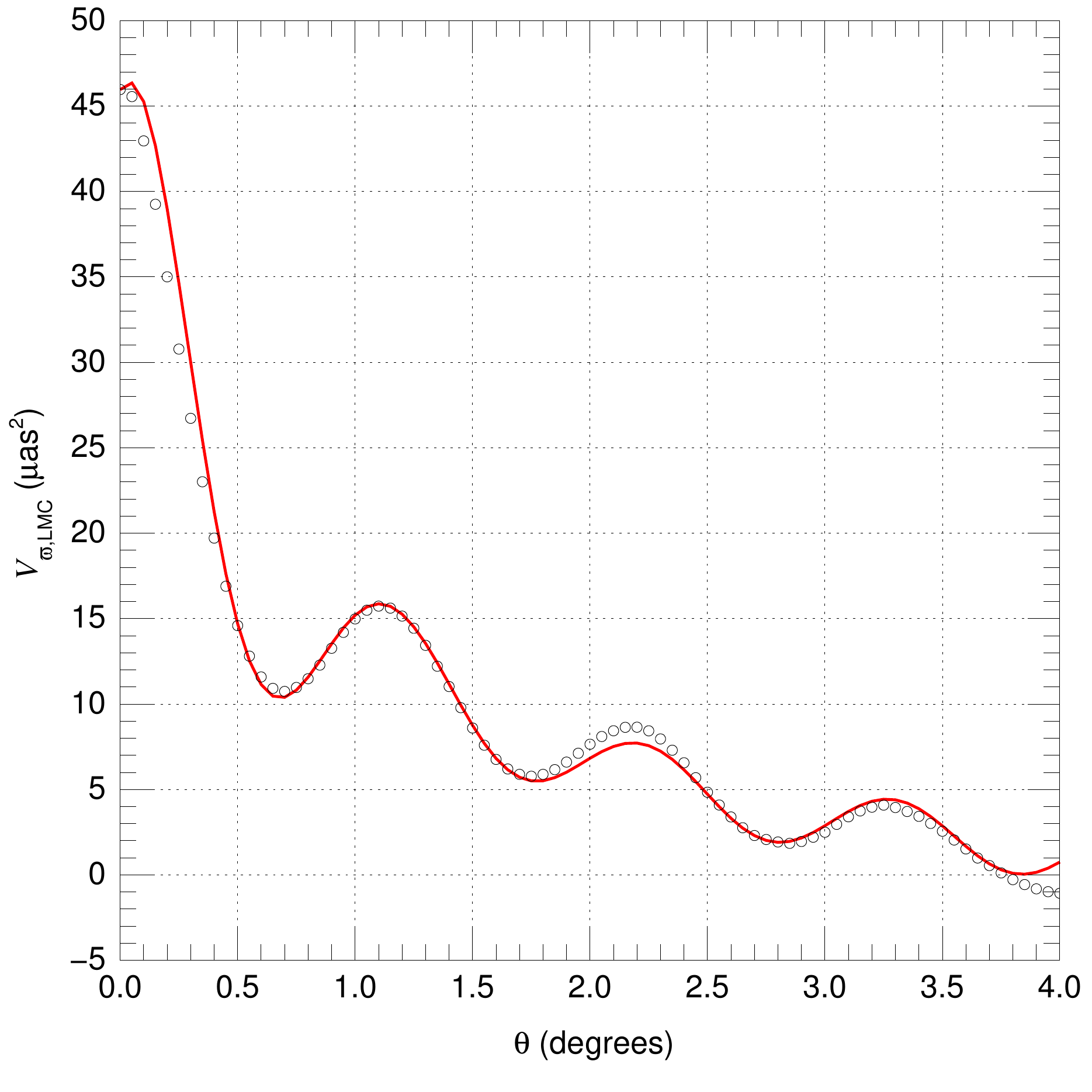}}
 \caption{(left) Smoothed corrected {\it Gaia}~EDR3 parallaxes for the bright LMC subsample. The axes are rectangular coordinates in degrees and the color table shows the 
          parallax scale in \microas. (right) Angular covariance in the LMC data with respect to the mean measured parallax.
          The points are the measured values and the red line is the analytical function described in the text.}
 \label{LMC_covar}
\end{figure*}

$\,\!$\indent To determine the angular covariance for small values of $\theta$ we use the {\it Gaia}~EDR3 LMC parallaxes to extend the work of L20b. We start by obtaining the sample
in a circular region with a radius of $10^{\rm o}$ centered at $\alpha = 81.28^{\rm o}$ and $\delta = -69.78^{\rm o}$ using the selection algorithm and coordinate transformation of
\citet{Lurietal20}. 
We obtain a bright subsample defined by $\GG < 19.019$, the median magnitude, to
study magnitude-dependent effects. We
apply \Zthree\ from L20a to calculate \pic\ for each star and calculate the weighted 
(from \sigmae)
average in the central $10^{\rm o}\times 10^{\rm o}$ region smoothed using a gaussian
kernel with a standard deviation of 0.1$^{\rm o}$. This is the same technique as L20b but with the differences of (a) using either the whole magnitude range or that with 
$\GG < 19.019$ as opposed to that with $\GG=16-18$ and (b) applying the L20a parallax correction beforehand. 

For the full sample the smoothed parallax is $24.2\pm 7.0$~\microas, which is 4.0~\microas\ larger than the 20.2~\microas\ value derived from the \citet{Pietetal19} distance but within one 
sigma\footnote{Note that 7.0~\microas\ is the dispersion of the smoothed values, not the standard deviation of the mean, which is much lower but irrelevant for the {\it Gaia}~EDR3 overall 
measurement for the LMC due to the second term in Eqn~\ref{spig}. Also, the average of the individual parallax values is different, as we discuss below, due to the non-uniform spatial coverage
of the sample.}
The equivalent result for the bright subsample is $24.7\pm 6.8$~\microas. The similarity between the full and bright samples indicates that the L20a correction is doing a good job 
of removing magnitude- and color-dependent effects in the $\GG = 18-20$~range
though we note that the dispersion for the faint subsample is 14.8~\microas, as it is dominated by the significantly larger individual uncertainties. Furthermore, the dispersion is nearly 
identical to the 6.9~\microas\ value found by L20b using the $\GG=16-18$ range,
indicating that
it has little magnitude dependence and that the application of the L20a correction does not introduce changes in the dispersion, as expected. Therefore, we conclude that the covariance 
at zero separation from the LMC data is well established to be $V_{\varpi,{\rm LMC}}(0) = 46.2$~\microas$^2$ within a few \microas$^2$, where we have used the value for our bright 
subsample.

The left panel of Fig.~\ref{LMC_covar} shows the smoothed parallax for the bright subsample (the full subsample is not shown but the equivalent plot is nearly identical), a plot 
equivalent to Fig.~14 in L20a. This pattern is commonly referred as the checkered pattern. We use those data to measure the angular covariance at 0.05$^{\rm o}$ intervals in the 
0$^{\,\rm o}$-4$^{\,\rm o}$ range. The results are shown in the right panel of Fig.~\ref{LMC_covar}. We also tried different analytical functions to use as an approximation to the 
fitted data and we settled on:

\begin{eqnarray}
 \label{LMC_equation}
 \frac{V_{\varpi,{\rm LMC}}}{V_{\varpi,{\rm LMC}}(0)} & = & a e^{-\theta/1.4} + \\
  &  & (1-a)\frac{\cos\left(2\pi\theta/\lambda+\phi\right)}{\cos\phi}\left(b e^{-(\theta/0.35)^{0.8}} + (1-b)\right), \nonumber
\end{eqnarray}

\noindent where $\theta$ is in degrees, $\lambda = 1.05^{\rm o}$, $a = 0.6$, $b = 0.94$, and $\phi = -5\pi/18$. Why fit an approximate analytical function? Two advantages are that
it is easier to implement than a tabulated form and that it allows for a better analysis of its behavior. However, one should be careful that no large discrepancies or anomalous
asymptotic behaviors are introduced. With respect to the first issue, we verified that in the 0$^{\,\rm o}$-4$^{\,\rm o}$ range the difference between the analytical function and the
data has a mean of 0.3~\microas$^2$ and a standard deviation of 1.0~\microas$^2$, which is sufficient for our needs. Regarding the asymptotic behavior at large separations, we generated
a numerical model with a periodic checkered pattern similar to the one seen in the left panel of Fig.~\ref{LMC_covar} (and even more clearly on the right panel of Fig.~14 in L20a) 
and an infinite extent and repeated the calculation up to separations of 40$^{\,\rm o}$.
The object of such a numerical model is to evaluate the periodic component of the checkered pattern to separate it from other contributions to the angular covariance.

\begin{figure*}
 \centerline{\includegraphics[width=0.49\linewidth]{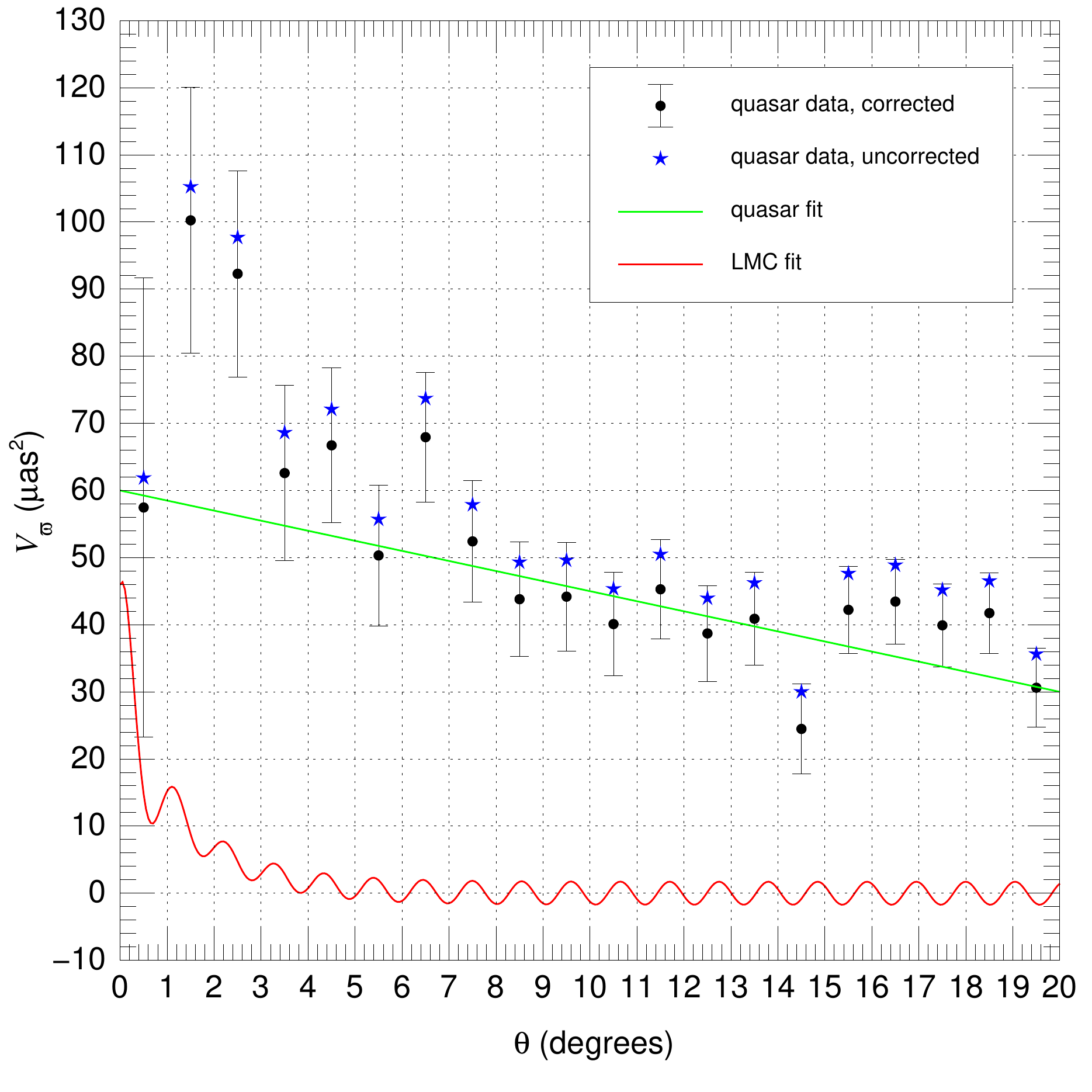} \
             \includegraphics[width=0.49\linewidth]{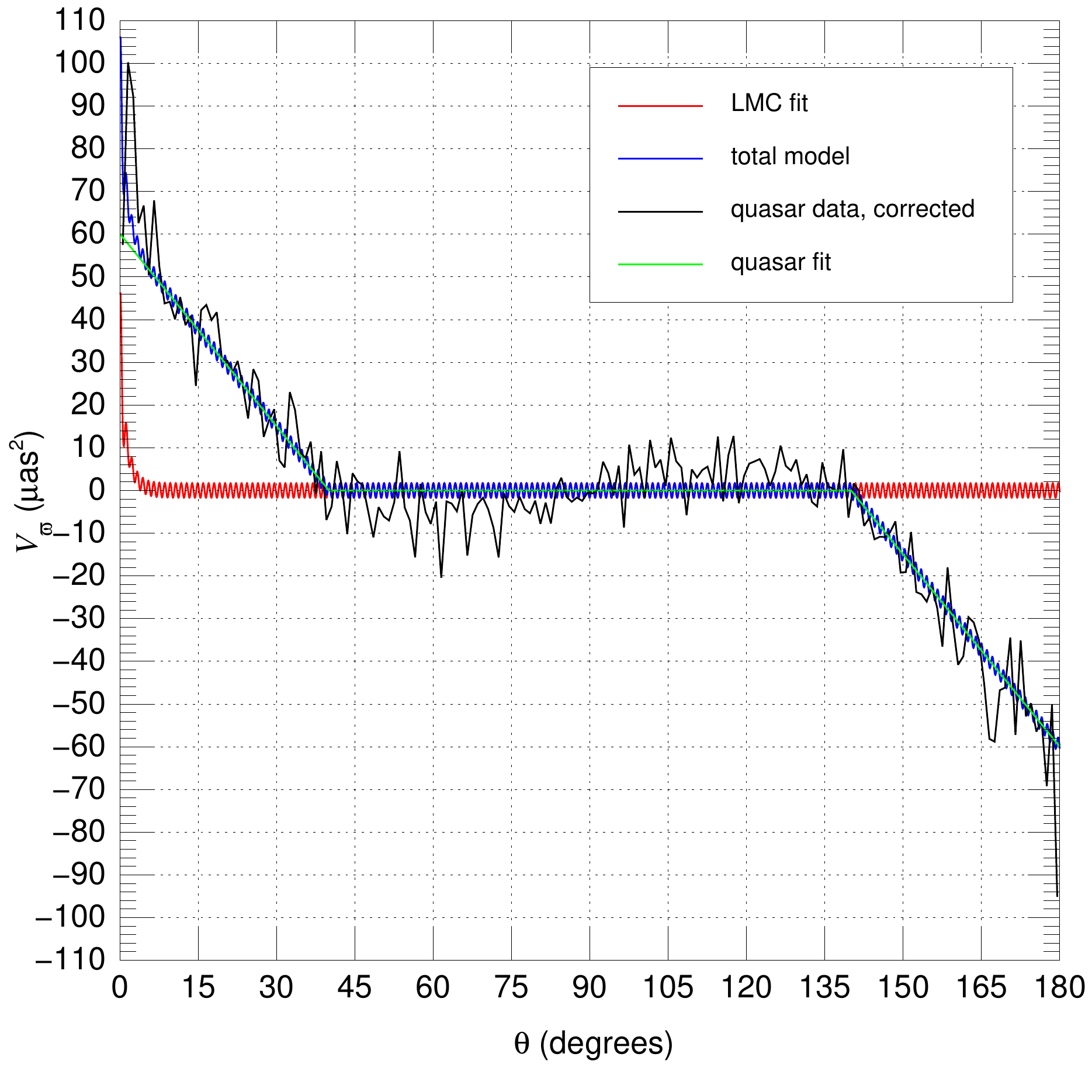}}
 \caption{Large-angle angular covariance of the {\it Gaia}~EDR3 parallaxes. The left panel shows the 0$^{\,\rm o}$-20$^{\,\rm o}$ separation range with the quasar 
          data in 1$^{\rm o}$ bins using either $\varpi$ 
          (blue asterisks)
          or \pic\ (black error bars), the linear fit to the quasar corrected data in the 
          5$^{\,\rm o}$-40$^{\,\rm o}$ separation range (green line), and the analytical fit to the LMC data (red line). The error bars associated to the quasar
          $\varpi$ data are not shown but they are similar to the \pic\ ones. The right panel shows the full 0$^{\,\rm o}$-180$^{\,\rm o}$ separation range and
          includes the total quasar+LMC model.}
 \label{full_covar}
\end{figure*}

The analytical function is the sum of an exponential and a damped sinusoid with the highest values at very small angles. The exponential component (first term in
Eqn.~\ref{LMC_equation}) indicates the existence of an uncorrected parallax bias that is correlated on scales of a few degrees. As we will see in the next subsection, such an effect 
has to be taken in conjunction with the angular covariance at larger separations. The damped sinusoid (second term in Eqn.~\ref{LMC_equation}) is the effect of the checkered pattern 
itself and its behavior is very well reproduced by the numerical model with infinite extent. The wavelength $\lambda$ as measured in the angular covariance is 1.05$^{\rm o}$ but note that
there is a $-5\pi/18$ phase in the cosine (the fitted function does not have a positive slope at zero due to the effect of the other terms). The oscillation is quickly damped but a 
small residual with a 2.4\% amplitude of the initial one ($\sim$~1~\microas$^2$) is maintained at large separations in the numerical model, where the phase is also conserved for at 
least several tens of cycles. This behavior is ultimately unphysical (the celestial sphere is finite and curved, among other reasons, so the pattern cannot be maintained) but given the 
small amplitude compared to other effects (see Fig.~\ref{full_covar} in the next subsection), its effect in our model is insignificant.

\subsection{Adding quasars for larger angles}

$\,\!$\indent To analyze the contribution from larger angles to the spatial covariance, we follow the same strategy of L20b but with two differences. First, we restrict the quasar
sample of L20b (\num{1214779} objects) to $\GG < 19$~mag and Galactic latitude $|b| > 25^{\rm o}$ to better simulate brighter objects and to minimize contamination effects close to the 
Galactic plane. This leaves us with a sample of \num{139036}~objects with a median \GG~=~18.58~mag. Second, we consider both uncorrected and corrected parallaxes.

We group the quasar pairs in 1$^{\rm o}$ bins and calculate the mean angular covariance in each bin. Results are shown in Fig.~\ref{full_covar}.

\begin{figure}
 \centerline{\includegraphics[width=\linewidth]{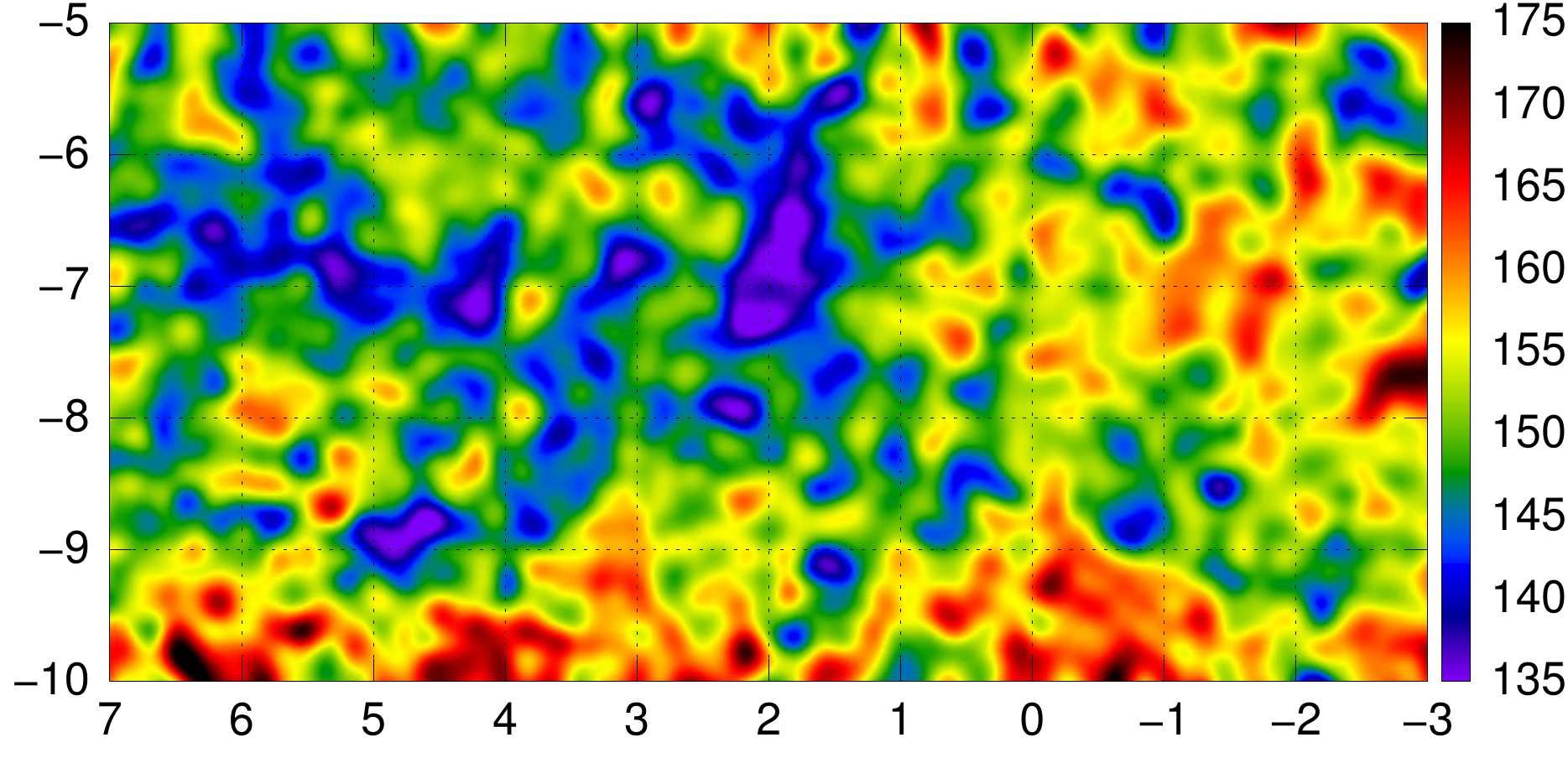}}
 \caption{Smoothed corrected {\it Gaia}~EDR3 parallaxes for the RC bulge sample. The axes are in Galactic coordinates in degrees and the color table shows the 
          parallax scale in \microas.}
 \label{bulge_covar}
\end{figure}

The qualitative behavior of the quasar covariance $V_{\varpi,{\rm QSO}}$ in Fig.~\ref{full_covar} is similar to that in Fig.~15 of L20b: a maximum at small separations, a nearly flat
regime for most values of $\theta$ and a minimum at large separations that is antisymmetric with respect to the one at small values. However, the amplitude at the extremes is 
significantly lower. The overall behavior using \pic\ is well fitted by the function:

\begin{equation}
 \begin{array}{llrcr}
  V_{\varpi,{\rm QSO}} = & 1.62(40-\theta)  &           3^{\,\rm o} \le & \theta \le &  40^{\,\rm o} \\
                         & 0                &          40^{\,\rm o} <   & \theta \le & 140^{\,\rm o} \\
                         & 1.62(140-\theta) & \;\;\;\;140^{\,\rm o} <   & \theta \le & 180^{\,\rm o} \\
 \end{array} ,
 \label{QSO_equation}
\end{equation}

\noindent with $\theta$ in degrees and $V_{\varpi,{\rm QSO}}$ in \microas$^2$, 
where for the linear fit we have used the 3$^{\,\rm o}$-40$^{\,\rm o}$ range. 
The value at 3$^{\rm o}$ (60~\microas$^2$) is substantially lower than the 142~\microas$^2$ value of L20b for zero separation.
This is a combination of three effects. Two are the use of 
a linear function instead of an exponential, which contributes to a reduction of $\sim$10~\microas$^2$, and the use of \pic\ instead of $\varpi$, which is a smaller effect of 
$\sim$5~\microas$^2$ (see Fig.~\ref{full_covar}). The third effect is the use of brighter quasars and that seems to be the dominant one. 

We obtain our final full covariance model by combining the results from the LMC and quasars from Eqns.~\ref{LMC_equation}~and~\ref{QSO_equation}:

\begin{equation}
 \angcov=V_{\varpi,{\rm LMC}}+V_{\varpi,{\rm QSO}},
 \label{fullcov}
\end{equation}

\noindent where for the 0-3$^{\rm o}$ range one should take $V_{\varpi,{\rm QSO}} = 60$~\microas$^2$ (Fig.~\ref{full_covar}). 
Note that the checkered pattern is also detected in the cumulative angular power
spectrum for quasars (Fig.~16 in L20b) and its effect is seen as deviations from the linear behavior at small separations in the left panel of
Fig.~\ref{full_covar} but we excluded those angles from the linear fit (where the error bars from the quasars are large in any case). 
Nevertheless, the sum of the two components in Eqn.~\ref{fullcov} is not far from what is seen for quasars at small separations.

\subsection{Limitations and validity}

$\,\!$\indent What are the limitations and validity of these results? There are 
two issues to analyze: the angular covariance for small scales and the overall angular covariance.

In order to test the validity of the angular covariance for small angles in other parts of the sky we chose the region of the Galactic bulge shown in Fig.~\ref{bulge_covar},
which was already used by \citet{Arenetal18} for the validation of the {\it Gaia}~DR2 parallaxes. We refined the exact location of the area in Galactic latitude by selecting a region not too
close to the Galactic plane (where extinction effects dominate) but not too far either (where there are not enough stars).
We selected the {\it Gaia}~EDR3 sources with Reduced Unit Weight Error RUWE~$<$~1.4 and 
\sigmai~$<$~0.1~mas and filtered the sample by choosing sources [a] within a proper motion radius of the expected center and [b] in 
the region of the \GBP$-$\GRP\ vs. \GG\ CMD that corresponds to red clump (RC) stars at the distance of the bulge (accounting for possible extinction). The results were then smoothed in
the same manner as for the LMC sample. The outcome, shown in Fig.~\ref{bulge_covar}, has a standard deviation of 6.7~\microas, which is remarkably similar to the LMC result,
and is much smaller than the equivalent value estimated from Fig.~13 in \citet{Arenetal18} for {\it Gaia}~DR2.
Furthermore, the sample has \GG\ magnitudes in the 14.5-17.6 range i.e. significantly brighter than the LMC sample. 

Two other analyses of the small-separation angular covariance for {\it Gaia}~EDR3 parallaxes became available while or shortly after this paper was submitted. \citet{VasiBaum21} uses 
globular clusters to obtain a value of $V_{\varpi}(0)$ around 50~\microas$^2$ that is very similar to ours. 
In the second paper, \citet{Zinn21} uses data from the Kepler field for this purpose.
In his Fig.~5, the first
data point for the angular covariance has a value of $\sim$45~$\microas^2$ but the trend with separation points towards a lower value around 16~$\microas^2$. In terms of the standard deviation,
the corresponding values are in the range of 4-7~\microas, similar to our results. Therefore, the value of $V_{\varpi}(0)$ as derived from regions of small angular size seems to be relatively 
independent of sky position and magnitude but we cannot discard that small changes are present.

As for larger separations,
given that the covariance from (mostly fainter) quasars in L20b is larger than the one here, it is possible that
the effect may be smaller for even brighter targets. However, the coherence between the LMC results for different magnitude ranges indicates that there is a limit 
to those possible improvements and,
as we have seen above,
it is also possible that the small-separation angular covariance may be slightly different in other parts of the sky from the LMC.  In summary, 
Eqn.~\ref{fullcov}
is a conservative estimate for the covariance in {\it Gaia}~EDR3 parallaxes but it may not be the
final word in the matter. Using that equation leads to a total \angcov(0)~=~106~\microas, 
with similar contributions from the LMC (small separations) and QSO (large separations) 
components, and to \sigmas~=~10.3~\microas\ as the proposed value to be used in \ref{sigmae} for {\it Gaia}~EDR3 parallaxes.

\section{Anchoring of the \textit{Gaia}~EDR3 parallaxes}

$\,\!$\indent Next, we briefly discuss the anchoring of the EDR3 parallaxes as a distance measurement.
Most distance measurements in astronomy are derived from photometric magnitudes. In that 
case, a constant uncertainty (random or systematic) in magnitude translates into a constant $\sigma_d/d$ i.e. relative distance uncertainties derived from magnitudes are (to first order)
independent of distance. Parallaxes are different, as to first order the uncertainty in distance derived from a parallax measurement grows as $d^2$ (ignoring issues with priors, see
\citealt{Maiz05c}). This means that for nearby objects parallactic distances will be in general more precise than photometric distances but for distant objects the situation will be reversed.
Therefore, from the calibration point of view we should use distant objects to calibrate parallaxes under the assumption that two objects in the same region of the sky and with the same magnitude 
and color will have the same parallax bias in {\it Gaia} (see below). 
In that way, if we have an object at infinity with an accurate (i.e. bias-free) parallax of zero, the relative parallax 
between the distant and the nearby object will effectively be an absolute parallax.

\begin{table}
\caption{Objects used to anchor the {\it Gaia}~EDR3 parallaxes. Uncertainties lower than 0.1~\microas\ are not listed.}
\centerline{
\begin{tabular}{lr@{$\pm$}lll}
\hline
Object(s)      & \mcii{\pig}       & \mci{Expected}           & Reference          \\
               & \mcii{(\microas)} & \mci{(\microas)}         &                    \\
\hline
bright quasars &    0.2&0.5        &           $<$0.1         & ---                \\
LMC            &   23.0&7.0        &         $\,$20.2$\pm$0.2 & \citet{Pietetal19} \\
SMC            &   20.1&7.5        &         $\,$15.9$\pm$0.6 & \citet{Cionetal00} \\
M31            &    9.8&10.1       & \phantom{$<$}1.3         & \citet{Connetal12} \\
M33            & $-$9.6&10.7       & \phantom{$<$}1.2         & \citet{Connetal12} \\
\hline
\label{anchors}                  
\end{tabular}
}
\end{table}

The approach described in the previous paragraph is the logic behind the L20a and L20b analysis: {\it Gaia}~EDR3 parallaxes are ultimately anchored in the values for quasars, which are forced to 
be zero through the use of parallax bias corrections. 
However, as discussed in the previous paragraph, parallaxes can also be anchored with galaxies for which accurate and precise distances exist because in those cases the uncertainties associated
with their photometric distances can be significantly lower than the uncertainties associated with their parallactic distances.
Here and in Table~\ref{anchors}
we present further lines of evidence that indicate that the {\it Gaia}~EDR3 parallaxes are correctly anchored:

\begin{enumerate}
 \item Could there be a magnitude dependency of the quasar EDR3 parallaxes? We have checked that by using the bright quasar subsample described in the previous subsection (as opposed to the 
       full sample used by L20a). 
       Their corrected mean parallax is within one sigma of zero (see Table~\ref{anchors}), where for the uncertainty we have simply used the standard deviation of the mean given that quasars
       are spread over the whole sky, so the answer to the question is no.
 \item What is the average EDR3 corrected parallax of the LMC? The values given above were calculated using a spatial smoothing
       in the region shown in Fig.~\ref{LMC_covar}. Doing a straight application of Eqns.~\ref{pig}~and~\ref{spig} in the region located within 10$^{\rm o}$ of the LMC center for
       the \citet{Lurietal20} sample
       instead yields the value in Table~\ref{anchors},
       with the second term in Eqn~\ref{spig} being the dominant contribution. This is just 0.4~sigmas above the expected value.
 \item We can ask the same question about the SMC. If we do the same application of Eqns.~\ref{pig}~and~\ref{spig} to the region located within 10$^{\rm o}$ of the SMC center for
       the \citet{Lurietal20} sample we obtain a (corrected) group parallax that is within 0.6~sigmas of the expected value.
 \item We also calculated the (corrected) group parallaxes for M31 and M33 using a similar procedure to the one described below for globular clusters. 
       For M31 we selected 1229 {\it Gaia}~EDR3 and for M33 we selected 2309 {\it Gaia}~EDR3 stars. The results in Table~\ref{anchors} 
       are within one sigma of the expected values, one above and one below zero, providing another indication of the absence of a significant
       bias in the EDR3 parallaxes.
\end{enumerate}

Therefore, we conclude that {\it Gaia}~EDR3 parallaxes are well anchored with respect to distant objects and can be considered bias-free in the sense that the values of a large sample 
located at an infinite distance and with a wide variety of sky positions, magnitudes, and colors will have an average corrected parallax close to zero. Problems may arise when we restrict the
sample in position, magnitude, and color and that is the subject of the next section.

\begin{table}
 \caption{Filters applied to the globular clusters and the reference SMC field in this paper used for the 47~Tuc analysis.
 \GG\ is filtered as $<$~18, RUWE as $<$~1.4 and \sigmae\ as $<$~0.1~mas in all cases.
 $\alpha$ and $\delta$ are the central right ascension and declination, respectively, and \pmra\ and \pmdec\ are their associated central proper motions.
 Note that the results for the SMC refer not to the galaxy as a whole but just to the region at the same position as 47~Tuc.
 }
\centerline{
\begin{tabular}{lccrccc}
\hline
ID           & $\alpha$ & $\delta$ & \mci{$r$}       & \pmra   & \pmdec   & $r_\mu$ \\
             & (deg)    & (deg)    & \mci{(\arcmin)} & (mas/a) & (mas/a)  & (mas/a) \\
\hline
$\omega$~Cen &   201.70 & $-$47.48 &              45 & $-$3.26 &  $-$6.74 &     1.0 \\
47~Tuc       &     6.02 & $-$72.08 &              45 & $-$5.25 &  $-$2.58 &     1.0 \\
SMC          &     6.02 & $-$72.08 &              45 & $+$0.51 &  $-$1.28 &     0.5 \\
NGC~6752     &   287.72 & $-$59.99 &              15 & $-$3.16 &  $-$4.04 &     1.0 \\
M5           &   229.64 &  $+$2.08 &              15 & $+$4.08 &  $-$9.86 &     1.0 \\
NGC~6397     &   265.17 & $-$53.68 &              20 & $+$3.24 & $-$17.65 &     1.0 \\
M13          &   250.42 & $+$36.46 &              15 & $-$3.13 &  $-$2.56 &     1.0 \\
\hline
\end{tabular}
}
\label{filters}                  
\end{table}

\begin{table*}
\caption{Globular cluster results for the parallax bias \Zfive\ and \kfive\ for five-parameter solutions.
         See the text for a description of the columns.}
\label{globularparallaxbias}
\centerline{
\begin{tabular}{rccrrrccc}
\mci{\GG\ range} & \nueff\ range & \nueff\ stats & \mci{$N$}   & \mci{$\Delta\varpi_{\rm const}$} & \mci{$\Delta\varpi_{\rm var}$} & $k_{5,\rm const}$ & $k_{5,\rm var}$ & $k_{5,\rm new}$ \\
\mci{(mag)}      & (\microninv)  & (\microninv)  &             & \mci{(\microas)}                 & \mci{(\microas)}               &                   &                 &                 \\
\hline
 9.3-11.0        & 1.25-1.60     & 1.37$\pm$0.06 & \num{76}    & $+14.4\pm2.2$                    &  $+6.9\pm2.2$                  & 1.19              & 1.10            & 1.11            \\
11.0-12.0        & 1.24-1.57     & 1.40$\pm$0.03 & \num{400}   & $+16.8\pm1.6$                    &  $+1.8\pm1.5$                  & 1.48              & 1.43            & 1.43            \\
12.0-13.0        & 1.30-1.72     & 1.45$\pm$0.04 & \num{847}   & $+15.7\pm1.3$                    &  $-2.8\pm1.3$                  & 1.81              & 1.69            & 1.70            \\
13.0-16.0        & 1.60-1.87     & 1.70$\pm$0.04 & \num{1454}  &  $-2.5\pm1.5$                    &  $+3.4\pm1.6$                  & 1.36              & 1.44            & 1.44            \\
13.0-16.0        & 1.40-1.60     & 1.50$\pm$0.02 & \num{7620}  &  $-4.2\pm0.5$                    &  $+1.3\pm0.5$                  & 1.35              & 1.32            & 1.32            \\
16.0-17.5        & 1.44-1.95     & 1.55$\pm$0.04 & \num{18703} &  $-1.6\pm0.6$                    &  $-3.0\pm0.6$                  & 1.25              & 1.21            & 1.21            \\
17.5-18.0        & 1.41-1.86     & 1.57$\pm$0.02 & \num{7421}  &  $-3.7\pm1.1$                    &  $-4.0\pm1.1$                  & 1.15              & 1.14            & 1.14            \\
\hline
\end{tabular}
}
\end{table*}

\section{Validating \textit{Gaia} EDR3 parallaxes with globular clusters}

$\,\!$\indent Once we have a model covariance for {\it Gaia}~EDR3 parallaxes, we can apply it to determine the distance uncertainties for stellar clusters. {\it Gaia} provides an
unparalleled combination of parallaxes, proper motions, and photometry that can be used to first select a clean sample of cluster members and then derive the distance to the system. 
For our analysis we follow a modification of the technique developed by \citet{Maiz19} 
and used for 16 stellar groups in \citet{Maizetal20b} and apply it to six globular clusters.
The results will be used in this section to validate {\it Gaia}~EDR3 parallaxes and in the next one to derive the distances to the globular clusters.
The selected globular clusters are the six richer known examples located far from the Galactic plane.
We use rich clusters to have a sample of stars as large and clean as possible. The
distance from the Galactic plane also minimizes contamination and differential extinction. The globular clusters and the filters used to select the sample 
are given in Table~\ref{filters}.
We use the results to test the parallax bias correction, derive the value for $k$, and characterize RUWE.

We download from the {\it Gaia}~EDR3 archive the sources in a $1^{\rm o}\times 1^{\rm o}$ or $2^{\rm o}\times 2^{\rm o}$ square (depending on the angular size of the object) centered on
each cluster and we select first the stars with $\GG < 18$, RUWE~$<$~1.4, and \sigmae~$<$~0.1~mas (for the latter we assume $k=1.0$, see below). We then apply a cut in 
angular distance
to the center of the cluster $r$ and an equivalent one in proper motion distance $r_\mu$. We do not apply a CMD cut as in the previous two papers, as 
our objects are the dominant population in the region and there is no significant differential extinction along their sightlines. We do the final 
selection by dropping the outliers in normalized parallax:

\begin{equation}
\pin = \frac{\pic-\pig}{\sigmae}
\label{pin}
\end{equation}

\noindent using a 4$\sigma$ cut. As explained in \citet{Maizetal20b}, the goal of this strategy is to have a final sample as clean as possible by sacrificing completeness. In other words, 
there must be many more stars in these clusters with {\it Gaia}~EDR3 entries.

\begin{figure}
 \centerline{\includegraphics[width=\linewidth]{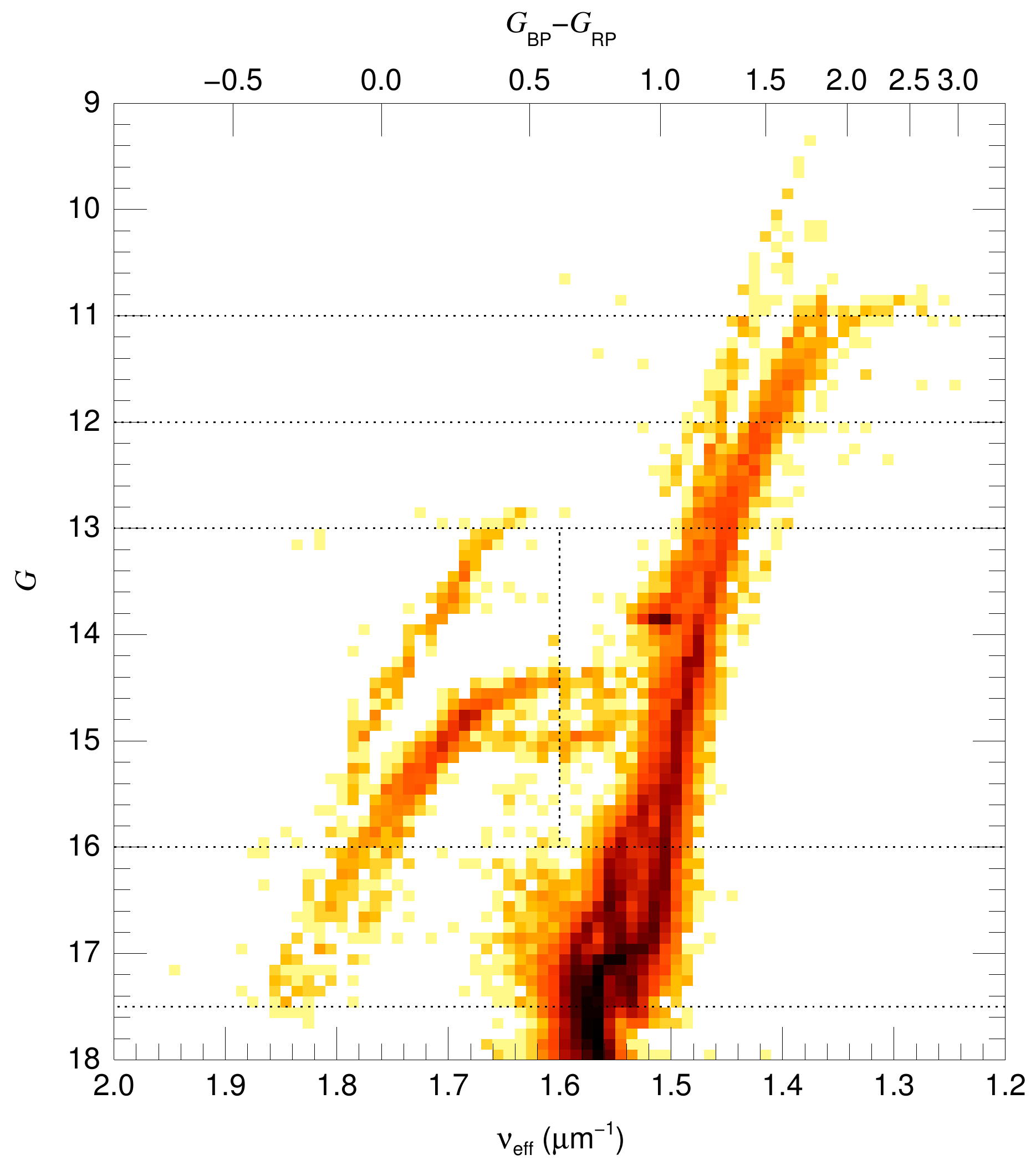}}
 \caption{Combined CMD for the six globular clusters using a logarithmic intensity scale. The dotted lines separate the CMD regions described in the text.}
 \label{CMD}
\end{figure}

\subsection{Testing the parallax bias correction as a function of \textit{G}}

$\,\!$\indent To test the parallax bias correction we first assume $k = 1.0$ in Eqn.~\ref{sigmae} and use either (a) a constant \Zthree\ value of $-$17~\microas\ or (b) the variable 
correction from Eqn.~\ref{Zdef}. We do the two selection procedures independently and obtain the cluster parallaxes and their uncertainties from Eqns.~\ref{pig}-\ref{spig}. We 
then select the stars with 5-parameter solutions (i.e. we only analyze \Zfive), subtract the corresponding cluster parallax from each star, and combine the resulting relative parallaxes 
$\Delta\varpi$ (star minus cluster) for all clusters, noting that the individual uncertainties for each star are larger than the spread expected from line-of-sight distance differences i.e. 
the relative parallaxes are almost exclusively caused by the random and systematic uncertainties and not by cluster-depth effects (see \citealt{Soltetal21}).
The CMD resulting from combining the six globular clusters is shown in Fig.~\ref{CMD}, where the 
conversion from \nueff\ to \GBP$-$\GRP\ is taken from Eqn.~2 in L20a. The CMD has the typical appearance of globular clusters, starting at the bottom from main-sequence (MS) stars, 
continuing to the top with red giants (RGs), and extending towards the left with horizontal branch (HB) stars, but, being a composite of six globular clusters, it shows multiple sequences.

As shown in Fig.~20 of L20a, for a given color and ecliptic latitude \Zfive\ is nearly-constant in a series of magnitude ranges and has abrupt changes at their boundaries. Therefore, to
validate \Zfive\ we divide our CMD using those magnitude ranges in Fig.~\ref{CMD} and Table~\ref{globularparallaxbias}. The next logical step would be to divide the CMD also in color 
ranges, as the changes in \Zfive\ as a function of \nueff\ are comparable to those as a function of \GG. Unfortunately, the CMD shows little spread in color for a given magnitude with the
only exception of the $G=13-16$ range, where most of the HB stars are found, so that is the only range where we do such a division. That leaves us with seven ranges and 
in each of those ranges we compute the mean and the standard deviation of $\Delta\varpi$ assuming either the constant \Zthree\ or the variable \Zfive\ (Table~\ref{globularparallaxbias}).

The comparison between $\Delta\varpi_{\rm const}$ and $\Delta\varpi_{\rm var}$ shows two clear magnitude ranges. For $\GG < 13$ the variable \Zfive\ is a significant improvement over 
the constant \Zthree, especially for $11 < \GG < 13$. On the other hand, for $\GG > 13$ differences are small: in all ranges $|\Delta\varpi|$ is at most 4.2~\microas, with the variable
\Zfive\ being better for $13 < \GG < 16$ and the constant \Zthree\ better for $16 < \GG < 18$. The advantage of $\Delta\varpi_{\rm var}$ for $\GG > 13$  is that it is sometimes positive 
and sometimes negative, leading to a small effect when combining magnitude ranges, while $\Delta\varpi_{\rm const}$ is always negative. In either case, $|\Delta\varpi|$ is significantly
smaller than \sigmae, which is the most relevant comparison. 

As it is clear that the variable \Zfive\ yields significant lower residuals overall when combining large magnitude ranges (especially for brighter stars), our main conclusion is that 
using it is recommended. However, we note that $\Delta\varpi_{\rm var}$ is relatively large for $\GG < 11$, that we have left blue stars mostly unstudied, and that our sampling of 
ecliptic latitudes is poor. A similar effect but with larger error bars is seen in Fig.~4 of \citet{Zinn21}.  Therefore, further testing is needed and that may lead to tweaking \Zfive.

\subsection{Deriving the values for \textit{k}}

$\,\!$\indent We now turn to the analysis of \kfive, the multiplicative constant in Eqn.~\ref{sigmae} for {\it Gaia}~EDR3 five-parameter solutions. We evaluate it in each of the ranges 
defined above by forcing the distribution in \pin\ in each one of them to have a standard deviation of 1.0. We do it by assuming either 
[a] the constant \Zthree\ ($k_{5,\rm const}$), [b] the variable \Zfive\ ($k_{5,\rm var}$), or [c] the variable \Zfive\ plus an additional correction introduced for all the stars in the 
range to force $\Delta\varpi$ to be zero ($k_{5,\rm new}$).
Results are listed in the three last columns of 
Table~\ref{globularparallaxbias}. A comparison between them shows that from $k_{5,\rm const}$ to $k_{5,\rm var}$ there is (a) a small reduction for bright stars, (b) a small increase 
for the HB range, and (c) little change for red faint stars. Between $k_{5,\rm var}$ and $k_{5,\rm new}$ differences are negligible. This indicates that the origin of $k$ being larger than 
one lies mostly in the underestimation of the random uncertainties, as it should, and not in the uncorrected parallax bias. If one wants to use a simple approximate formula to implement
$k$, one possibility is:

\begin{equation}
 \begin{array}{llrlr}
  \kfive = & 1.1             &      & G \le & 11 \\
           & 1.1 + 0.6(G-11) & 11 < & G \le & 12 \\
           & 1.7 - 0.1(G-12) & 12 < & G \le & 18 \\
           & 1.1             & 18 < & G          \\
 \end{array} ,
 \label{k_equation}
\end{equation}

\noindent which, we note, is quite similar to the results found in Fig.~19 of \citet{Fabretal20} for $\GG > 12$ but significantly lower for bright stars. The same structure as a 
function of \GG\ and similar values for $k$ are seen in Fig.~16 of \citet{ElBaetal21}.

\begin{table}
\caption{Globular cluster results as a function of RUWE for five-parameter solutions.
         See the text for a description of the columns.}
\centerline{
\begin{tabular}{lrrccc}
\hline
RUWE    & $N_{*,0}$   & $N_*$       & $\langle\pin\rangle$ & \sigman & $k_{5,\rm ext}$ \\
\hline
0.0-1.4 & \num{30689} & \num{30592} &  $-0.01$             &  $0.99$ &            1.00 \\
1.4-2.0 & \num{550}   & \num{548}   &  $-0.01$             &  $1.00$ &            1.66 \\
2.0-3.0 & \num{63}    & \num{63}    &  $-0.13$             &  $1.01$ &            1.92 \\
\hline
\end{tabular}
}
\label{RUWE}                     
\end{table}

\begin{table*}
\caption{Membership and astrometric results for the globular clusters and the reference SMC field in this paper used for the 47~Tuc analysis.
$N_{*,0}$ and $N_*$ are the number of objects before and after the
normalized parallax 4-sigma cut; $t_\varpi$, $t_{\mu_{\alpha *}}$, $t_{\mu_{\delta}}$ are reduced-$\chi^2$-like statistics for those three quantities; \pig, \pmrag, and \pmdecg are
the derived astrometric values; and $d$ is the distance after applying a flat prior truncated at 20~kpc.}
\centerline{
\renewcommand{\arraystretch}{1.2}
\begin{tabular}{lrrcccr@{$\pm$}lr@{$\pm$}lr@{$\pm$}lr@{}lll}
\hline
ID                & $N_{*,0}$   & $N_*$       & \tpi  & \tra & \tdec & \mcii{\pig}       & \mcii{\pmrag}  & \mcii{\pmdecg}  & \mcii{$d$}              & \mci{$d_{\rm lit}$}    & Ref. \\
                  &             &             &       &      &       & \mcii{(\microas)} & \mcii{(mas/a)} & \mcii{(mas/a)}  & \mcii{(kpc)}            & \mci{(kpc)}            &      \\
\hline
$\omega$~Cen      & \num{6865}  & \num{6814}  &  1.02 & 6.73 &  6.22 & 192.8&9.6         & $-$3.258&0.020 &  $-$6.749&0.020 & 5.25&$^{+0.28}_{-0.25}$ & 5.2                    & H10  \\
                  &             &             &       &      &       & \mcii{}           & \mcii{}        & \mcii{}         & \mcii{}                 & 5.19$^{+0.07}_{-0.08}$ & W15  \\
                  &             &             &       &      &       & \mcii{}           & \mcii{}        & \mcii{}         & \mcii{}                 & 5.44                   & C20  \\
                  &             &             &       &      &       & \mcii{}           & \mcii{}        & \mcii{}         & \mcii{}                 & 5.24$\pm$0.11          & S21  \\
47~Tuc            & \num{15350} & \num{15298} &  1.03 & 4.74 &  4.76 & 227.8&9.7         & $+$5.260&0.020 &  $-$2.567&0.020 & 4.42&$^{+0.20}_{-0.18}$ & 4.81$\pm$0.18          & R05  \\
                  &             &             &       &      &       & \mcii{}           & \mcii{}        & \mcii{}         & \mcii{}                 & 4.5                    & H10  \\
                  &             &             &       &      &       & \mcii{}           & \mcii{}        & \mcii{}         & \mcii{}                 & 4.15$\pm$0.08          & W15  \\
                  &             &             &       &      &       & \mcii{}           & \mcii{}        & \mcii{}         & \mcii{}                 & 4.45$\pm$0.12          & C18  \\ 
                  &             &             &       &      &       & \mcii{}           & \mcii{}        & \mcii{}         & \mcii{}                 & 4.55$\pm$0.03          & T20  \\
                  &             &             &       &      &       & \mcii{}           & \mcii{}        & \mcii{}         & \mcii{}                 & 4.50$\pm$0.07          & T20  \\
                  &             &             &       &      &       & \mcii{}           & \mcii{}        & \mcii{}         & \mcii{}                 & 4.53$\pm$0.06          & TW   \\
SMC               & \num{1904}  & \num{1900}  &  0.93 & 1.51 &  1.38 &  22.8&9.0         & $+$0.506&0.018 &  $-$1.275&0.018 & \mcii{---}              & ---                    & ---  \\
NGC~6752          & \num{2793}  & \num{2784}  &  1.00 & 4.01 &  4.15 & 256.5&10.2        & $-$3.175&0.022 &  $-$4.044&0.022 & 3.92&$^{+0.16}_{-0.15}$ & 4.0                    & H10  \\
                  &             &             &       &      &       & \mcii{}           & \mcii{}        & \mcii{}         & \mcii{}                 & 4.02$^{+0.10}_{-0.08}$ & W15  \\
                  &             &             &       &      &       & \mcii{}           & \mcii{}        & \mcii{}         & \mcii{}                 & 4.06                   & C20  \\
M5                & \num{1210}  & \num{1206}  &  0.94 & 2.74 &  2.64 & 137.3&10.4        & $+$4.083&0.022 &  $-$9.863&0.022 & 7.46&$^{+0.64}_{-0.54}$ & 8.28$\pm$0.31          & R05  \\
                  &             &             &       &      &       & \mcii{}           & \mcii{}        & \mcii{}         & \mcii{}                 & 7.5                    & H10  \\
                  &             &             &       &      &       & \mcii{}           & \mcii{}        & \mcii{}         & \mcii{}                 & 7.79$^{+0.47}_{-0.61}$ & W15  \\
                  &             &             &       &      &       & \mcii{}           & \mcii{}        & \mcii{}         & \mcii{}                 & 7.62                   & C20  \\
NGC~6397          & \num{4090}  & \num{4082}  &  0.93 & 4.40 &  4.83 & 418.5&10.1        & $+$3.239&0.022 & $-$17.649&0.022 & 2.40&$^{+0.06}_{-0.06}$ & 2.3                    & H10  \\
                  &             &             &       &      &       & \mcii{}           & \mcii{}        & \mcii{}         & \mcii{}                 & 2.39$^{+0.13}_{-0.11}$ & W15  \\
                  &             &             &       &      &       & \mcii{}           & \mcii{}        & \mcii{}         & \mcii{}                 & 2.40                   & C20  \\
M13               & \num{1850}  & \num{1848}  &  1.06 & 3.06 &  2.98 & 128.1&10.3        & $-$3.128&0.022 &  $-$2.564&0.022 & 8.02&$^{+0.73}_{-0.62}$ & 7.98$\pm$0.29          & R05  \\
                  &             &             &       &      &       & \mcii{}           & \mcii{}        & \mcii{}         & \mcii{}                 & 7.1                    & H10  \\
                  &             &             &       &      &       & \mcii{}           & \mcii{}        & \mcii{}         & \mcii{}                 & 7.24                   & C20  \\
\hline
{\it References.} & \multicolumn{15}{l}{C18: \citet{Chenetal18b}. C20: \citet{Cernetal20}. H10: \citet{Harr10}. R05: \citet{Recietal05}. S21: \citet{Soltetal21}.} \\
                  & \multicolumn{15}{l}{T20: \citet{Thometal20}. TW: This work using the SMC as reference.  W15: \citet{Watketal15}.}
\end{tabular}
\renewcommand{\arraystretch}{1.0}
}
\label{results}                  
\end{table*}

\subsection{How bad is a bad RUWE?}             

$\,\!$\indent RUWE is the goodness-of-fit statistic recommended by L20b as the main filtering criterion to exclude stars with poor astrometric solutions. It is an adimensional 
quantity with an average of $\sim$1.0 and the most common value used for filtering is 1.4 (as we have done above for the globular clusters in this paper). However, if a star with
RUWE of, say, 1.3 is considered to be have a good astrometric solution, a valid question would be: how bad is the solution for another one with, say, 1.5? And what if the RUWE is 2.5? 
Here we do a simple analysis to quantify that.

As described in our procedure above, the last step in the selection of the sample for each globular cluster is a cut in \pin\ where we drop the 4$\sigma$ outliers. Following
\citet{Maizetal20b}, we define the numbers in the sample before and after the cut as $N_{*,0}$ and $N_*$, respectively. Their values are given in Table~\ref{RUWE} for the whole sample
considering only stars with five-parameter solutions and in Table~\ref{results} cluster by cluster for stars with five- and six-parameter solutions. After the application of $k$, the final 
sample should have a \pin\ distribution with an average close to zero and a standard deviation \sigman\ close to one and that is indeed the case for RUWE~$<$~1.4 in Table~\ref{RUWE}.

To evaluate the degradation of the quality of the parallaxes with increasing RUWE we include in our analysis the five-parameter stars selected by our algorithm but excluded solely on the
basis of their RUWE. We divide them in two RUWE ranges (see Table~\ref{RUWE}), introduce an additional multiplicative factor $k_{5,\rm ext}$ on top of the one from Eqn.~\ref{k_equation}, and 
increase it from one until we force the standard deviation of \pin\ to be $\sim$1
(and, hence, $\sigman\sim 1$).
When we do that, we find that for RUWE between 1.4 and 2.0 (a) less than 1\% of the points are farther away
than 4$\sigma$, (b) the required value for $k_{5,\rm ext}$ is moderately small and (c) the resulting distribution has an average close to zero. This indicates that those parallaxes are likely 
safe to use after introducing the extra uncertainty. For RUWE between 2.0 and 3.0, on the other hand, $k_{5,\rm ext}$ is already close to two, the average \pin\ has started to deviate from zero
(very slightly),
and our sample is small. Therefore, those values may still be used but the possibility of them being biased may be larger.

\section{\textit{Gaia}~EDR3 distances to globular clusters and going beyond the angular covariance limit}

$\,\!$\indent The results for the globular clusters are given in Table~\ref{results}. To calculate the distance we used a flat prior truncated at 20~kpc, so no a priori knowledge about the 
spatial distribution of globular clusters is used. Nevertheless, given the small relative uncertainties for \pig\ (all better than 10\%), the results are robust. For example, changing the
distance at which the prior is truncated to 15~kpc or 30~kpc leaves the results unchanged (see Fig.~1 in \citealt{Maiz05c}).

We also list in Table~\ref{results} literature results for the globular clusters in the sample. There is a general good agreement, which is especially significative in some cases. For the
two closest clusters, NGC~6397 and NGC~6752, the discrepancies are very small, at the level of 0.1~kpc or less. For 47~Tuc there is a very good agreement with the high-precision results
from \citet{Thometal20} using eclipsing binaries.

A special situation is that of the \citet{Soltetal21} result for $\omega$~Cen using the same {\it Gaia}~EDR3 parallaxes as us. Their distance is essentially the same as ours (difference of
0.01~kpc, within the rounding error) but their uncertainty is significantly smaller, as their result for \pig\ is 191$\pm$1~(stat.)~$\pm$4~(syst.)~\microas, where the statistical uncertainty
should be understood to be that from the first term in our Eqn.~\ref{spig} and the systematic one that of the second term. The authors claim that the 4~\microas\ is derived from the L20b
analysis of the LMC but, as we have seen in this paper, that value is actually 6.8~\microas\ and when the effect of terms from larger separations is included it grows to 10.3~\microas. The
reduction to 9.6~\microas\ in Table~\ref{results} is achieved by the averaging effect introduced by the considerable size of the cluster (Fig.~\ref{full_covar}).

Is there a way to beat the {\it Gaia}~EDR3 angular covariance limit of $\sim$10~\microas? In principle, it can be done if one has a collection of additional sources with an external 
high-precision distance measurement and the same spatial distribution as the cluster one is interested in. Those additional sources can be used as a reference to trace the uncorrected 
{\it Gaia}~EDR3 parallax bias. Indeed, that is the principle behind using quasars to study the spatial covariance, as their true parallaxes are of the order of nanoarcseconds. Could quasars
be used as a reference for a particular cluster? Unfortunately, no. There are few known quasars per square degree and their individual {\it Gaia}~EDR3 uncertainties are too large to be
useful.

A background galaxy such as one of the MCs is a different story. As one of our clusters, 47~Tuc, is notoriously located in front of the SMC, it can be used to see if it is possible to beat
the angular covariance limit. 47~Tuc and the SMC are well differentiated in proper motions, so it is easy to separate the components from the two systems in exactly the same region. That is
what we did here using the filters in Table~\ref{filters} and processing the {\it Gaia}~EDR3 parallaxes for the SMC population behind in 47~Tuc to derive the SMC results listed in 
Table~\ref{results}. The distance to the SMC given by \citet{Cionetal00} is 62.8$\pm$0.8~(stat.)~$\pm$2.3~(syst.)~kpc. That corresponds to a parallax of 15.9$\pm$0.6~\microas. The value we
derive from {\it Gaia}~EDR3 parallaxes is 22.8$\pm$9.0~\microas\ but that includes both the random and systematic components\footnote{The parallax is similar but not identical to the one
in Table~\ref{anchors}, which is calculated from the whole SMC, as expected given the effect of the angular covariance at separations of a few degrees.}.
The latter is not needed if we want to calculate relative 
parallaxes in the same region and with a similar spatial distribution. The random component for the SMC is just 1.5~\microas. Therefore, the uncorrected parallax bias is 6.9$\pm$1.6~\microas,
including in the error budget that of the SMC distance. The value for the \pic\ uncertainty for 47~Tuc in Table~\ref{results} also includes the systematic component; eliminating it
leaves an uncertainty of just 0.4~\microas. That leads to a final \pic\ for 47~Tuc corrected from the SMC parallax bias of 220.9$\pm$1.7~\microas, where the uncertainty is purely statistical. 
However, we should remember the results of Table~\ref{globularparallaxbias} and that the SMC sample is significantly fainter than the 47~Tuc, leading to a possible magnitude-dependent bias. 
A more realistic uncertainty including those effects is 3~\microas, which yields a final distance of 4.53$\pm$0.06~kpc. Therefore, for 47~Tuc we are able to beat the angular covariance limit
and derive a precise distance that is in excellent agreement with the eclipsing-binary result of \citet{Thometal20}.
The result is also within one sigma of the {\it Gaia}~DR2 value by \citet{Chenetal18b}, who used a similar (but not identical) method to the one we used here and whose uncertainty is twice 
as large as ours (a good example of the improvement from DR2 to EDR3).



\section{Summary and recipe for using \textit{Gaia} EDR3 parallaxes}

$\,\!$\indent In this paper we have presented a procedure to obtain accurate and precise {\it Gaia}~EDR3 parallaxes. The procedure includes the correction of the known parallax biases and
the addition of the unknown biases into the error budget. To do that correctly, we have performed an analysis of the angular covariance of the {\it Gaia}~EDR3 parallaxes combining LMC and 
quasar data and verified it with Galactic bulge data. A sample of six globular clusters has been used to validate the procedure and, while doing so, we have derived accurate distances to 
them. For 47~Tuc we have shown that one can go beyond the angular covariance limit
with the help of the background stars in the SMC.

The biases related to angular covariance are likely to be hard to eliminate. The hope there is that DR4 will significantly reduce them, as EDR3 did for the ones present in DR2. However, it
may be possible to reduce the magnitude/color/position biases that are also remaining in the data without having to wait for DR4, especially for the brightest stars. We plan to do so in the
near future by combining the results in this paper with an analysis of open clusters that may lead to an improvement of \Zfive.

We wrap up this paper by giving a recipe for the derivation of {\it Gaia}~EDR3 distances based on the analysis in this paper:

\begin{enumerate}
 \item Do a preliminary filtering by RUWE, five- or six-parameter solutions, and other criteria (e.g. parallax uncertainty) to eliminate objects with undesirable properties (this
       depends on the specific task at hand).
 \item Apply the known parallax bias correction using Eqns.~\ref{pic}~and~\ref{Zdef} here and the information in the L20a tables.
 \item Convert from internal to external parallax uncertainties using Eqn.~\ref{sigmae} with \sigmas~=~10.3~\microas\ and $k$ from Eqn.~\ref{k_equation} with the corrections from
       Table~\ref{RUWE} if using objects with RUWE in the 1.4-3.0 range.
 \item If one is combining two or more parallaxes, apply Eqns.~\ref{pig},~\ref{weights},~and~\ref{spig} with 
       the angular covariance taken from Eqns.~\ref{LMC_equation},~\ref{QSO_equation},~and~\ref{fullcov}.
 \item Select an appropriate prior and calculate the posterior distribution for the distance.
 \item If background/foreground targets of a known distance are present in the field, consider using them to beat the angular covariance limit.
\end{enumerate}

\begin{acknowledgements}
The authors thank Lennart Lindegren and the rest of the DPAC astrometry team for their extraordinary work in producing {\it Gaia}~EDR3 and
Stefano Casertano and Adam Riess for useful discussions on the topics of this paper.
We also thank Xavier Luri for letting us have access to the LMC {\it Gaia}~EDR3 selection script.
J.M.A. and M.P.G. acknowledge support from the Spanish Government Ministerio de Ciencia through grant PGC2018-\num{095049}-B-C22. 
R.H.B. acknowledges support from the ESAC visitors program.
This work has made use of data from the European Space Agency (ESA) mission {\it Gaia} ({\tt https://www.cosmos.esa.int/gaia}), 
processed by the {\it Gaia} Data Processing and Analysis Consortium (DPAC, {\tt https://www.cosmos.esa.int/web/gaia/dpac/consortium}).
Funding for the DPAC has been provided by national institutions, in particular the institutions participating in the {\it Gaia} Multilateral Agreement. 
This research has made use of the SIMBAD and VizieR databases, operated at CDS, Strasbourg, France. 
\end{acknowledgements}


%

\bibliographystyle{aa} 
\bibliography{general} 

\end{document}